# Gravitino Production in the Inflationary Universe and the Effects on Big-Bang Nucleosynthesis

M. Kawasaki

*Institute for Cosmic Ray Research, The University of Tokyo, Tanashi 188, Japan*

T. Moroi [*]

*Department of Physics, Tohoku University, Sendai 980, Japan*

(March 29, 1994)

## Abstract

Gravitino production and decay in the inflationary universe are reexanimed. Assuming that gravitino mainly decays into photon and photino, we have calculated the upperbound of the reheating temperature. Compared to previous works, we have essentially improved the following two points; (i) the helicity $\pm\frac{3}{2}$ gravitino production cross sections are calculated by using the full relevant terms in the supergravity lagrangian, and (ii) high energy photon spectrum is obtained by solving the Boltzmann equations numerically. Photo-dissociation of the light elements (D, T, $^3$He, $^4$He) leads to the most stringent upperbound of the reheating temperature, which is given by $(10^6$–$10^9)$GeV for the gravitino mass 100GeV–1TeV. On the other hand, requiring that the present mass density of photino should be smaller than the critical density, we find that the reheating temperature have to be smaller than $(10^{11}$–$10^{12})$GeV for the photino mass (10–100)GeV, irrespectively of the gravitino mass. The effect of other decay channel is also considered.

Typeset using REVTEX

---

[*]Fellow of the Japan Society for the Promotion of Science.

# I. INTRODUCTION

When one thinks of new physics beyond the standard model, supersymmetry (SUSY) is one of the most attractive candidates. Cancellation of quadratic divergences in SUSY models naturally explains the stability of the electroweak scale against radiative corrections [1,2]. Furthermore, if we assume the particle contents of the minimal SUSY standard model (MSSM), the three gauge coupling constants in the standard model meet at $\sim 2 \times 10^{16}$ GeV [3,4], which strongly supports the grand unified theory (GUT).

In spite of these strong motivations, no direct evidence for SUSY (especially superpartners) has been discovered yet. This means that the SUSY is broken in nature, if it exists. Although many efforts have been made to understand the origin of the SUSY breaking, we have not understood it yet. Nowadays, many people expect the existence of *local* SUSY (i.e., supergravity) and try to break it spontaneously in this framework. In the broken phase of the supergravity, super-Higgs effect occurs and gravitino, which is the superpartner of graviton, acquires mass by absorbing the Nambu-Goldstone fermion associated with the SUSY breaking sector. In many broken local SUSY models, we expect that the mass of gravitino $m_{3/2}$ is at the same order of those of squarks and sleptons since the following (tree level) super-trace formula among the mass matrices $\mathcal{M}_J^2$'s holds [5];

$$\mathrm{Str}\mathcal{M}^2 \equiv \sum_{\mathrm{spin}\,J} (-1)^{2J}(2J+1)\mathrm{tr}\mathcal{M}_J^2 \simeq 2(n-1)m_{3/2}^2, \qquad (1)$$

where $n$ is the number of chiral multiplet. For example in models with the minimal kinetic term, this is the case and all the SUSY breaking masses of squarks and sleptons are equal to gravitino mass at the Planck scale. But contrary to our theoretical interests, we have no hope to see gravitinos directly in the collider experiments since the interaction of gravitino is extremely weak.

On the other hand, if we assume the standard big-bang cosmology, the mass of gravitino is severely constrained. If gravitino is unstable, it may decay after the big-bang nucleosynthesis (BBN) and produces an unacceptable amount of entropy, which conflicts with the predictions of BBN. In order to keep success of BBN, the gravitino mass should be larger than $\sim 10$ TeV as Weinberg first pointed out [6]. Meanwhile, in the case of stable gravitino, its mass should be smaller than $\sim 1$ keV not to overclose the universe [7]. Therefore, the gravitino mass between $\sim 1$ keV and $\sim 10$ TeV conflicts with the standard big-bang cosmology.

However, if the universe went through inflation, we may avoid the above constraints [8] since the initial abundance of gravitino is diluted by the exponential expansion of the universe. But even if the initial gravitinos are diluted, the above problems still potentially exist since gravitinos are reproduced by scattering processes off the thermal radiation after the universe has been reheated [9–17]. The number density of secondary gravitino is proportional to the reheating temperature and hence, upperbound on the reheating temperature should be imposed not to overproduce gravitinos. Therefore, even assuming the inflation, a detailed analysis must be made to obtain the upperbound on the reheating temperature. The case of stable gravitino is analyzed in Refs. [13,15,17] and we will not deal with it.

In this paper, we assume that the gravitinos are unstable and derive the upperbound on the reheating temperature. The analysis of the cosmological regeneration and decay of unstable gravitino has been done in many articles. These previous works show that the most stringent upperbound on the reheating temperature comes from the photo-dissociation of the light nuclei (D, T, $^3$He, $^4$He). Once gravitinos are produced in the early universe, most of them decay after BBN since the lifetime of gravitino with mass $O(100\text{GeV} - 10\text{TeV})$ is $O((10^8 - 10^2)\text{sec})$. If gravitinos decay radiatively, emitted high energy photons induce cascade processes and affect the result of BBN. Not to change the abundance of light nuclei, we must constrain the number density of gravitinos, and this constraint is translated into the upperbound of reheating temperature.

In order to analyze the photo-dissociation process, we must calculate the following two quantities precisely; the number density of gravitinos produced after the reheated universe, and the high energy photon spectrum induced by radiative decay of gravitinos. But the previous estimations of these values are incomplete. As for the number density of gravitino, most of the previous works follow the result of Ref. [11], where the number density is underestimated by factor ∼4. Furthermore, in many articles, the spectrum of high energy photon, which determines the photo-dissociation rates of light elements, are calculated by using a simple fitting formula. In this paper, we have treated these effects precisely and found the more stringent upperbound on the reheating temperature than the previous calculations. The plan of this paper is as follows. In Sec.II, we calculate the gravitino production cross section in the early universe. In Sec.III, high energy photon spectrum induced by the radiative decay of gravitino is obtained by solving the Boltzmann equation numerically. The results are shown in Sec.V. Other cosmological constraints is considered in Sec.VI and Sec.VII is devoted to discussions.

## II. GRAVITINO PRODUCTION IN THE EARLY UNIVERSE

After the universe has reheated, gravitinos are reproduced by the scattering processes of the thermal radiations and decay with decay rate of order of $m_{3/2}^3/M_{pl}^2$ where $M_{pl} = \sqrt{8\pi}M \sim 1.22 \times 10^{19}$GeV is the Planck mass. Since the interaction of gravitino is very weak, gravitino cannot be thermalized if the reheating temperature $T_R$ is less than $O(M_{pl})$. In this case, Boltzmann equation for the gravitino number density $n_{3/2}$ can be written as

$$\frac{dn_{3/2}}{dt} + 3Hn_{3/2} = \langle \Sigma_{tot} v_{rel} \rangle n_{rad}^2 - \frac{m_{3/2}}{\langle E_{3/2} \rangle} \frac{n_{3/2}}{\tau_{3/2}}, \qquad (2)$$

where $H$ is the Hubble constant, $\Sigma_{tot}$ is the total cross section which is derived in appendix A, $\langle \cdots \rangle$ means thermal average, $n_{rad} \equiv \zeta(3)T^3/\pi^2$ represents the number density of the scalar boson in thermal bath, $v_{rel}$ is the relative velocity of the scattering radiations ($\langle v_{rel} \rangle = 1$ in our case), and $m_{3/2}/\langle E_{3/2} \rangle$ is the averaged Lorenz factor. Note that the first term of right hand side (r.h.s.) of Eq.(2) represents contribution from the gravitino production process, and the second one comes from the decay of gravitino. In Eq.(2), we have omitted the terms which represents the inverse processes since their contributions are unimportant at low temperature that we are interested in. For the radiation dominated universe, $H$ is given by

$$H \equiv \frac{\dot{R}}{R} = \sqrt{\frac{N_* \pi^2}{90 M^2}}\, T^2, \qquad (3)$$

where $R$ is the scale factor and $N_*$ is the total number of effectively massless degrees of freedom. For the particle content of MSSM, $N_*(T_R) \sim 228.75$ if $T_R$ is much larger than the masses of the superpartners, and $N_*(T \ll 1\text{MeV}) \sim 3.36$.

At the time right after the end of the reheating of the universe, the first term dominates the r.h.s. of Eq.(2) since gravitinos have been diluted by the de Sitter expansion of the universe. Using yield variable $Y_{3/2} \equiv n_{3/2}/n_{rad}$ and ignoring the decay contributions, Eq.(2) becomes

$$\frac{dY_{3/2}}{dT} = -\frac{\langle \Sigma_{tot} v_{rel} \rangle n_{rad}}{HT}, \tag{4}$$

where we have assumed the relation

$$RT = \text{const}. \tag{5}$$

Ignoring the small $T$-dependence of $\Sigma_{tot}$, we can solve Eq.(4) analytically. Integrating Eq.(4) from the reheating temperature $T_R$ to $T$ ($T_R \gg T$) and multiplying the dilution factor $N_S(T)/N_S(T_R)$, the yield of gravitino is found to be

$$Y_{3/2}(T) = \frac{N_S(T)}{N_S(T_R)} \times \frac{n_{rad}(T_R) \langle \Sigma_{tot} v_{rel} \rangle}{H(T_R)}. \tag{6}$$

For the MSSM particle content, $N_S(T_R) \sim 228.75$ and $N_S(T \ll 1\text{MeV}) \sim 3.91$. Eq.(6) shows that $Y_{3/2}$ is proportional to $T_R$. Note that the numerical value of $Y_{3/2}$ in our case is about $4-5$ times larger than the result in Ref. [11]. Some comments on this difference are given in section VII. From Eq.(6), we can derive the simple fitting formula for $Y_{3/2}$;

$$Y_{3/2}(T \ll 1\text{MeV}) \simeq 2.14 \times 10^{-11} \left(\frac{T_R}{10^{10}\text{GeV}}\right) \left\{1 - 0.0232 \log\left(\frac{T_R}{10^{10}\text{GeV}}\right)\right\}, \tag{7}$$

where the logarithmic correction term comes from renormalization group flow of the gauge coupling constants. The difference between the exact formula (6) and the above approximated one is within $\sim 5\%$ ($\sim 25\%$) for $10^6$ GeV $\lesssim T_R \lesssim 10^{14}$ GeV ($10^2$ GeV $\lesssim T_R \lesssim 10^{19}$ GeV).

As the temperature of the universe drops and $H^{-1}$ approaches $\tau_{3/2}$, decay term becomes the dominant part of the r.h.s. of Eq.(2). Ignoring the scattering term, Eq.(2) can be rewritten as

$$\frac{dY_{3/2}}{dt} = -\frac{Y_{3/2}}{\tau_{3/2}}, \tag{8}$$

where we have taken $m_{3/2}/\langle E_{3/2} \rangle = 1$ since gravitinos are almost at rest. Using Eq.(6) as a boundary condition, we can solve Eq.(8) and the answer is

$$Y_{3/2}(t) = \frac{n_{3/2}(t)}{n_{rad}(t)} = \frac{N_S(T)}{N_S(T_R)} \times \frac{n_{rad}(T_R) \langle \Sigma_{tot} v_{rel} \rangle}{H(T_R)} \exp\left(-\frac{t}{\tau_{3/2}}\right), \tag{9}$$

where the relation between $t$ and $T$ can be obtained by solving Eq.(3) with Eq.(5);

$$t = \frac{1}{2}\sqrt{\frac{90M^2}{N_*\pi^2}} T^{-2}. \tag{10}$$

## III. RADIATIVE DECAY OF GRAVITINO

Radiative decay of gravitino may affect BBN. We analyze this effect assuming that gravitinos $\psi_\mu$ mainly decay to photons $\gamma$ and photinos $\tilde{\gamma}$.

In order to investigate the photo-dissociation processes, we must know the spectra of the high energy photon and electron induced by the gravitino decay. In this section, we will derive these spectra by solving the Boltzmann equations numerically.

Once high energy photons are emitted in the gravitino decay, they induce cascade processes. In order to analyze these processes, we have taken the following radiative processes into account. (I) High energy photon with energy $\epsilon_\gamma$ can become $e^+ e^-$ pair by scattering off background photon if the energy of the background photon is larger than $m_e^2/\epsilon_\gamma$ ( $m_e$: electron mass ). We call this process double photon pair creation. For sufficiently high energy photons, this is the dominant process since the cross section or the number density of target is much larger than other processes. Numerical calculation shows that this process determines the shape of the spectrum of high energy photon for $\epsilon_\gamma \gtrsim m_e^2/22T$. (II) Below the effective threshold of double photon pair creation, high energy photons lose their energy by photon-photon scattering. But in the limit of $\epsilon_\gamma \to 0$, the total cross section for the photon-photon scattering is proportional to $\epsilon_\gamma^3$ and this process loses its significance. Hence finally, photons are thermalized by (III) pair creation in nuclei, or (IV) Compton scattering off thermal electron. And (V) emitted high energy electrons and positrons lose their energy by the inverse Compton scattering off background photon. Furthermore, (VI) the source of these cascade processes are the high energy photons emitted in the decay of gravitinos. Note that we only consider the decay channel $\psi_\mu \to \gamma + \tilde{\gamma}$ and hence energy of the incoming photon $\epsilon_{\gamma 0}$ is monoclomatic.

The Boltzmann equations for the photon and electron distribution function $f_\gamma$ and $f_e$ are given by

$$\frac{\partial f_\gamma(\epsilon_\gamma)}{\partial t} = \left.\frac{\partial f_\gamma(\epsilon_\gamma)}{\partial t}\right|_{\mathrm{DP}} + \left.\frac{\partial f_\gamma(\epsilon_\gamma)}{\partial t}\right|_{\mathrm{PP}} + \left.\frac{\partial f_\gamma(\epsilon_\gamma)}{\partial t}\right|_{\mathrm{PC}}$$
$$+ \left.\frac{\partial f_\gamma(\epsilon_\gamma)}{\partial t}\right|_{\mathrm{CS}} + \left.\frac{\partial f_\gamma(\epsilon_\gamma)}{\partial t}\right|_{\mathrm{IC}} + \left.\frac{\partial f_\gamma(\epsilon_\gamma)}{\partial t}\right|_{\mathrm{DE}}, \tag{11}$$

$$\frac{\partial f_e(E_e)}{\partial t} = \left.\frac{\partial f_e(E_e)}{\partial t}\right|_{\mathrm{DP}} + \left.\frac{\partial f_e(E_e)}{\partial t}\right|_{\mathrm{PC}} + \left.\frac{\partial f_e(E_e)}{\partial t}\right|_{\mathrm{CS}} + \left.\frac{\partial f_e(E_e)}{\partial t}\right|_{\mathrm{IC}}, \tag{12}$$

where DP (PP, PC, CS, IC, and DE) represents double photon pair creation (photon-photon scattering, pair creation in nuclei, Compton scattering, inverse Compton scattering, and the contribution from the gravitino decay). Full details are shown in appendix B.

In order to see the photon spectrum, we have to solve Eq.(11) and Eq.(12). Since the decay rate of gravitino is much smaller than the scattering rates of other processes, gravitinos can be regarded as a stationary source of high energy photon at each moment. Therefore, we only need a stationary solution of Eq.(11) and Eq.(12) with non-zero $(\partial f_\gamma/\partial t)|_{\mathrm{DE}}$ at each temperature. Note that Eq.(11) and Eq.(12) are linear equations of $f_\gamma$ and $f_e$, and hence, once Eq.(11) and Eq.(12) have been solved with some reference value of $(\partial \tilde{f}_\gamma/\partial t)|_{\mathrm{DE}}$ we can reconstruct the photon spectrum for arbitrary value of $(\partial f_\gamma/\partial t)|_{\mathrm{DE}}$ with $T$ and $\epsilon_{\gamma 0}$ fixed;

$$f_\gamma(\epsilon_\gamma) = \tilde{f}_\gamma(\epsilon_\gamma) \times \frac{(\partial f_\gamma/\partial t)|_{\rm DE}}{(\partial \tilde{f}_\gamma/\partial t)|_{\rm DE}}. \qquad (13)$$

For each $T$ and $\epsilon_{\gamma 0}$, we have calculated the reference spectra $\tilde{f}_\gamma(\epsilon_\gamma)$ and $\tilde{f}_e(E_e)$ by solving Eq.(11) and Eq.(12) numerically with the condition,

$$\frac{\partial f_\gamma(\epsilon_\gamma)}{\partial t} = \frac{\partial f_e(E_e)}{\partial t} = 0. \qquad (14)$$

Typical spectra are shown in Figs.1 in which we have shown the case with $\epsilon_{\gamma 0} = 100{\rm GeV}$ and $10{\rm TeV}$, $T = 100{\rm keV}, 1{\rm keV}, 10{\rm eV}$, and the incoming flux of the high energy photon is normalized to be

$$\epsilon_{\gamma 0} \times \left.\frac{\partial \tilde{f}_\gamma(\epsilon_\gamma)}{\partial t}\right|_{\rm DE} = \delta(\epsilon_\gamma - \epsilon_{\gamma 0})\ {\rm GeV}^5. \qquad (15)$$

The behaviors of the photon spectra can be understood in the following way. In the region $\epsilon_\gamma \gtrsim m_e^2/22T$, the photon number density is extremely suppressed since the rate of double photon pair creation process is very large. Just below this threshold value, the shape of the photon spectrum is determined by the photon-photon scattering process, and if the photon energy is sufficiently small, the Compton scattering with the thermal electron is the dominant process for photons. In Fig.2, we have compared our photon spectrum with the results of the simple fitting formula used in Ellis et al [16]. As one can see, not only the absolute value but also the form of the spectrum differs between them. The fitting formula in Ref. [16] is derived from the numerical results given in Refs. [18,19] in which, however, the effect of the Compton scattering is not taken into account. Our results indicate that the number of Compton scattering events is comparable to that of the inverse Compton events for such low energy region, since the number density of the high energy electron is extremely smaller than that of high energy photon. Therefore, the deformation of the photon spectrum by Compton scattering is expected below the threshold of the photon-photon scattering.

## IV. BBN AND PHOTO-DISSOCIATION OF LIGHT ELEMENTS

BBN is one of great successes of the standard big bang cosmology. It is believed that light elements of mass number less than 7 are produced at early stage of the universe when the cosmic temperature is between 1MeV and 10keV. Theoretical predictions for abundances of light elements is excellently in good agreement with those expected from observations if baryon-to-photon ratio $\eta_B$ is about $3 \times 10^{-10}$.

However the presence of gravitino might destroy this success of BBN. Gravitino may have three effects on BBN. First the energy density of gravitino at $T \simeq 1{\rm MeV}$ speeds up the cosmic expansion and leads to increase the $n/p$ ratio and hence $^4$He abundance also increases. Second, the radiative decay of gravitino reduces baryon-to–photon ratio and results in too baryon-poor universe. Third, the high energy photons emitted in the decay of gravitino destroy the light elements. Among three effects, photo-dissociation by high energy photon is the most important for gravitino of mass less than $\sim 1{\rm TeV}$. In the following we consider the photo-dissociation of light elements and discuss other effects in Sec.VI.

The high energy photons emitted in the decay of gravitinos lose their energy during multiple electromagnetic processes described in the previous section. Surviving soft photons can destroy the light elements (D, T, $^3$He, $^4$He) if their energy are greater than the threshold of the photo-dissociation reactions. We consider the photo-dissociation reactions listed in Table I. For the process D($\gamma$,n)p, we have used the cross section in analytic form which is given in Ref. [21], and the cross sections for other reactions are taken from the experimental data (for references, see Table I). We neglect $^4$He($\gamma$, D)D and $^4$He($\gamma$, 2p 2n) since their cross sections are small compared with the other reactions. Furthermore, we do not include the photo-dissociation process for $^7$Li and $^7$Be because the cross section data for $^7$Be is not available and hence we cannot predict the abundance of $^7$Li a part of which come from $^7$Be.

The time evolution of the light elements are described by

$$\frac{dn_{\rm D}}{dt} = -n_{\rm D} \sum_i \int_{E_i} d\epsilon_\gamma \sigma^i_{{\rm D}\to a}(\epsilon_\gamma) f_\gamma(\epsilon_\gamma) + \sum_i \int_{E_i} d\epsilon_\gamma \sigma^i_{a\to {\rm D}}(\epsilon_\gamma) n_a f_\gamma(\epsilon_\gamma), \tag{16}$$

$$\frac{dn_{\rm T}}{dt} = -n_{\rm T} \sum_i \int_{E_i} d\epsilon_\gamma \sigma^i_{{\rm T}\to a}(\epsilon_\gamma) f_\gamma(\epsilon_\gamma) + \sum_i \int_{E_i} d\epsilon_\gamma \sigma^i_{a\to {\rm T}}(\epsilon_\gamma) n_a f_\gamma(\epsilon_\gamma), \tag{17}$$

$$\frac{dn_{^3{\rm He}}}{dt} = -n_{^3{\rm He}} \sum_i \int_{E_i} d\epsilon_\gamma \sigma^i_{^3{\rm He}\to a}(\epsilon_\gamma) f_\gamma(\epsilon_\gamma) + \sum_i \int_{E_i} d\epsilon_\gamma \sigma^i_{a\to ^3{\rm He}}(\epsilon_\gamma) n_a f_\gamma(\epsilon_\gamma), \tag{18}$$

$$\frac{dn_{^4{\rm He}}}{dt} = -n_{^4{\rm He}} \sum_i \int_{E_i} d\epsilon_\gamma \sigma^i_{^4{\rm He}\to a}(\epsilon_\gamma) f_\gamma(\epsilon_\gamma) + \sum_i \int_{E_i} d\epsilon_\gamma \sigma^i_{a\to ^4{\rm He}}(\epsilon_\gamma) n_a f_\gamma(\epsilon_\gamma), \tag{19}$$

where $\sigma^i_{a\to b}$ is the cross section of the photo-dissociation process $i$: $a+\gamma \to b+\ldots$ and $E_i$ is the threshold energy of reaction $i$. When the energy of the high energy photon is relatively low, i.e. 2MeV $\lesssim \epsilon_\gamma \lesssim$ 20MeV the D, T and $^3$He are destroyed and their abundances decrease. On the other hand, if the photons have high energy enough to destroy $^4$He, it seems that such high energy photons only decrease the abundance of all light elements. However since D, T and $^3$He are produced by the photo-dissociation of $^4$He whose abundance is much higher than the other elements, their abundances can increase or decrease depending on the number density of high energy photon. When the number density of high energy photons with energy greater than $\sim$ 20MeV is extremely high, all light elements are destroyed. But as the photon density becomes lower, there is some range of the high energy photon density at which the overproduction of D, T and $^3$He becomes significant. And if the density is sufficiently low, high energy photon does not affect the BBN at all.

From various observations, the primordial abundances of light elements are estimated [20] as

$$0.22 < Y_p \equiv \left(\frac{\rho_{^4{\rm He}}}{\rho_B}\right)_p < 0.24, \tag{20}$$

$$\left(\frac{n_{\rm D}}{n_{\rm H}}\right)_p > 1.8 \times 10^{-5}, \tag{21}$$

$$\left(\frac{n_{\rm D} + n_{^3{\rm He}}}{n_{\rm H}}\right)_p < 1.0 \times 10^{-4}, \tag{22}$$

where $\rho_{^4{\rm He}}$ and $\rho_B$ are the mass densities of $^4$He and baryon. The abundances of light elements modified by gravitino decay must satisfy the observational constraints above. In order to make precise predictions for the abundances of light elements, the evolutional equations (16) – (19) should be incorporated with the nuclear network calculation of BBN. Therefore, we have modified Kawano's computer code [28] to include the photo-dissociation processes.

From the above arguments it is clear that there are at least three free parameters, i.e. mass of gravitino $m_{3/2}$, reheating temperature $T_R$ and baryon-to-photon ratio $\eta_B$. Furthermore we also study the case in which gravitino has other decay channels. In the present paper we do not specify other decay channel. Instead, we introduce another free parameter $B_\gamma$ which is the branching ratio for the channel $\psi_\mu \to \gamma + \tilde{\gamma}$. Therefore we must study the effect of gravitino decay on BBN in four dimensional parameter space. However in the next section it will be shown that the baryon-to-photon ratio $\eta_B$ is not important parameter in the present calculation because the allowed value for $\eta_B$ is almost the same as that in the standard case (i.e. without gravitino).

## V. RESULTS

### A. $B_\gamma = 1$ case

First we have investigated the photo-dissociation effect when all gravitinos decay into photons and photinos ($B_\gamma = 1$). We have taken the range of three free parameters as $10\text{GeV} \leq m_{3/2} \leq 10\text{TeV}$, $10^5\text{GeV} \leq T_R \leq 10^{13}\text{GeV}$ and $10^{-10} \leq \eta_B \leq 10^{-9}$. In this calculation, we assume that photino is massless. The contours for the critical abundances of the light elements D, (D+$^3$He) and $^4$He in the $\eta_B - T_R$ plane are shown in Figs.3 for (a)$m_{3/2} = 10\text{GeV}$, (b)$100\text{GeV}$, (c)$1\text{TeV}$ and (d)$10\text{TeV}$, respectively. For low reheating temperature ($T_R \lesssim 10^6\text{GeV}$), the number density of the gravitino is very low and hence the number density of the induced high energy photons is too low to affect the BBN. Therefore the resultant abundances of light elements are the same as those in the standard BBN. The effect of the photo-dissociation due to gravitino decay becomes significant as the reheating temperature increases.

As seen in Figs.3, the allowed range of baryon-to-photon ratio is almost same as that without gravitino for $m_{3/2} \lesssim 1\text{TeV}$, i.e. very narrow range around $\eta_B \sim 3 \times 10^{-10}$ is allowed. However for $m_{3/2} \sim 1\text{TeV}$ and $T_R \sim 10^9\text{GeV}$ or $m_{3/2} \sim 1\text{TeV}$ and $T_R \sim 10^{12}\text{GeV}$, lower values of $\eta_B$ are allowed (Fig.3(c)). In this case, the critical photon energy ($\sim m_e^2/22T$) for double photon pair creation process is lower than the threshold of photo-dissociation reaction of $^4$He. Therefore, for $T_R \lesssim 10^{12}\text{GeV}$, the abundance of $^4$He is not affected by the gravitino decay. Then the emitted photons only destroy $^3$He and D whose abundances would be larger than the observational constraints for low baryon density if gravitino did not exist. Therefore one sees the narrow allowed band at $T = 10^9\text{GeV}$ where only a small number of $^3$He and D are destroyed to satisfy the constraints (21) and (22). For $T_R \gtrsim 10^{12}\text{GeV}$, since a large number of high energy photons are produced even above the threshold of double photon pair creation, a part of $^4$He are destroyed to produce $^3$He and D, which leads to the very narrow allowed region at $T_R \sim 10^{12}\text{GeV}$. However even in this special case, the upper limit of allowed reheating temperature is changed very little between $\eta_B = 10^{-10}$ and $\eta_B \sim 3 \times 10^{-10}$. This allows us to fix $\eta_B = 3.0 \times 10^{-10}$ in deriving the upperbound of the reheating temperature.

The allowed regions that satisfy the observational constraints (20)-(22) also shown in Figs.4 in the $m_{3/2} - T_R$ plane for $\eta_B = 3 \times 10^{-10}$. In Figs.3 and Fig.4(a) one can see four typical cases depending on $T_R$ and $m_{3/2}$.

- $m_{3/2} \lesssim 1\text{TeV}$, $T_R \lesssim 10^{11}\text{GeV}$:
  In this case the lifetime of the gravitino is so long that the critical energy for double photon process ($\sim m_e^2/22T$) is higher than the threshold of the photo-dissociation reactions for $^4$He at the decay time of gravitino. Thus $^4$He is destroyed to produce T, $^3$He and D. (Since T becomes $^3$He by $\beta$-decay, hereafter we mean T and $^3$He by the word "$^3$He".) Since the reheating temperature is not so high, the number density of gravitino is not high enough to destroy all the light elements completely. As a result, $^3$He and D are produced too much and the abundance of $^4$He decreases. To avoid the overproduction of $^3$He and D, the reheating temperature should be less than $\sim (10^6 - 10^9)\text{GeV}$.

- $m_{3/2} \lesssim 1\text{TeV}$, $T_R \gtrsim 10^{11}\text{GeV}$:
  The lifetime is long enough to destroy $^4$He and the gravitino abundance is so large that all the light elements are destroyed since the reheating temperature is extremely high. This parameter region is strongly excluded by the observation.

- $1\text{TeV} \lesssim m_{3/2} \lesssim 3\text{TeV}$:
  The lifetime becomes shorter as the mass of gravitino increases, and the decay occurs when double photon pair creation process works well. If the cosmic temperature at $t = \tau_{3/2}$ is greater than $\sim m_e^2/22E_{^4\text{He}}$ (where $E_{^4\text{He}} \sim 20\text{MeV}$ represents the typical threshold energy of of $^4$He destruction processes), $^4$He abundance is almost unaffected by the high energy photons as can be seen in Fig.3(c). In this parameter region, overproduction of (D+$^3$He) cannot occur since $^4$He is not destroyed. In this case, the destruction of D is the most important to set the limit of the reheating temperature. This gives the constraint of $T_R \lesssim 10^9 - 10^{12}\text{GeV}$.

- $m_{3/2} \gtrsim 3\text{TeV}$:
  In this case the decay occurs so early that all high energy photons are quickly thermalized by double photon process before they destroy the light elements. Therefore the effect on BBN is negligible. Fig.3(d) is an example of this case. The resultant contours for abundances of light elements are almost identical as those without the decay of gravitino.

### B. $B_\gamma < 1$ case

So far we have assumed that all gravitinos decay into photons and photinos. But if other superpartners are lighter than gravitino, the decay channels of gravitino increases and the branching ratio for the channel $\psi_\mu \to \gamma + \tilde{\gamma}$ becomes less than 1. In this case, various decay products affect the evolution of the universe and BBN. In this paper, instead of studying all decay channels, we consider only the $\gamma + \tilde{\gamma}$ channel with taking the branching ratio $B_\gamma$ as another free parameter. With this simplification, the effect of all possible decay products other than photon is not taken into account. Therefore the resultant constraints on the reheating temperature and the mass of gravitino should be taken as the conservative constraints since other decay products may destroy more light elements and make the constraints more stringent.

Although we have four free parameters in the present case, the result for $B_\gamma = 1$ implies that the allowed range of $T_R$ and $m_{3/2}$ is obtained if we take baryon-to-photon ratio to be $3 \times 10^{-10}$. Since our main concern is to set the constraints on $T_R$ and $m_{3/2}$, we can safely fix $\eta_B$ ($= 3 \times 10^{-10}$).

The constraints for $B_\gamma = 0.1$ and $B_\gamma = 0.01$ is shown in Fig.4(b) and Fig.4(c) which should be compared with Fig.4(a) ($B_\gamma = 1$ case). Since the number density of the high energy photons is proportional to $B_\gamma$, the constraint on the reheating temperature becomes less stringent as $B_\gamma$ decreases. In addition, the total lifetime of gravitino is given by

$$\tau_{3/2} = \tau(\psi_\mu \to \gamma + \tilde{\gamma}) \times B_\gamma. \tag{23}$$

Thus the gravitinos decay earlier than that for $B_\gamma = 1$ case and the constraints from ($^3$He + D) overproduction becomes less stringent. This effect can be seen in Fig.4(b), where the constraint due to the overproduction of ($^3$He + D) has a cut at $m_{3/2} \simeq 400\text{GeV}$ compared with $\sim 1\text{TeV}$ for $B_\gamma = 1$.

In Fig.5, the contours for the upperbound of reheating temperature are shown in the $m_{3/2} - B_\gamma$ plane. One can see that the stringent constraint on $T_R$ is imposed for $m_{3/2} \lesssim 100\text{GeV}$ even if the branching ratio is small. As mentioned before this constraint should be regarded as the conservative one and the actual constraint may become more stringent by the effect of other decay products, which will be investigated elsewhere.

## VI. OTHER CONSTRAINTS

In the previous section, we have considered the constraints from the photo-dissociation of light elements. But as we have seen, if the mass of gravitino is larger than a few TeV, gravitino decay does not induce light element photo-dissociation and no constraints has been obtained. In the case of such a large gravitino mass, we must consider other effects of gravitino.

If we consider the present mass density of photinos produced by the gravitino decay, we can get the upperbound of the reheating temperature. In SUSY models with $R$-invariance (which is the usual assumption), the lightest superparticle (in our case, photino) is stable. Thus the photinos produced by the decay of gravitinos survive until today, and they contribute to the energy density of the present universe. Since one gravitino produces one photino, we can get the present number density of the photino:

$$n_{\tilde{\gamma}} = Y_{3/2}(T \ll 1\text{MeV}) \times \frac{\zeta(3)}{\pi^2} T_0^3, \tag{24}$$

where $T_0$ is the present temperature of the universe. Density parameter of photino $\Omega_{\tilde{\gamma}} \equiv m_{\tilde{\gamma}} n_{\tilde{\gamma}}/\rho_c$ can be easily calculated, where $m_{\tilde{\gamma}}$ is the photino mass, $\rho_c \simeq 8.1 \times 10^{-47} h^2 \text{GeV}^4$ is the critical density of the universe and $h$ is the Hubble parameter in units of 100km/sec/Mpc. If we constrain that $\Omega_{\tilde{\gamma}} \leq 1$ in order not to overclose the universe, the upperbound of the reheating temperature is given by

$$T_R \leq 2.7 \times 10^{11} \left(\frac{m_{\tilde{\gamma}}}{100\text{GeV}}\right)^{-1} h^2 \text{ GeV}, \qquad (25)$$

where we have ignored the logarithmic correction term of $\Sigma_{tot}$. To set the upperbound on the reheating temperature, we need to know the mass of photino. If one assumes the gaugino-mass unification condition, the lower limit on the mass of photino is 18.4GeV [29]. Then we can get the following upperbound on the reheating temperature:

$$T_R \leq 7.1 \times 10^{11} h^2 \text{GeV}. \qquad (26)$$

Note that this bound is independent of the gravitino mass and branching ratio.

Another important constraint comes from the effect on the cosmic expansion at BBN. As mentioned before, if the density of gravitino at nucleosynthesis epochs becomes high, the expansion of the universe increases, which leads to more abundance of $^4$He. We study this effect by using modified Kawano's code and show the result in Fig.6. In the calculation, we take $\eta_B = 2.8 \times 10^{-10}$ and $\tau_n = 887\,\text{sec}$ (where $\tau_n = (889 \pm 2.1)$sec is the neutron lifetime [34]) so that the the predicted $^4$He abundance is minimized without conflicting the observational constraints for other light elements. The resultant upperbound of the reheating temperature is given by

$$T_R \lesssim 2 \times 10^{13} \text{GeV} \left(\frac{m_{3/2}}{1\text{TeV}}\right)^{-1}, \qquad (27)$$

for $m_{3/2} > 1\text{TeV}$.

## VII. DISCUSSIONS

We have investigated the production and the decay of gravitino, in particular, the effect on BBN by high energy photons produced in the decay. We have found that the stringent constraints on reheating temperature and mass of gravitino.

Let us compare our result with those in other literatures. Our number density of gravitino produced in the reheating epochs of the inflationary universe is about four times larger than that in Ellis et al [11]. Since Ellis et al [11] note that they have neglected the interaction terms between gravitino and chiral multiplets (which is the second term in Eq.(A1)), they might underestimate the total cross section for the production of gravitino. All previous works concerning gravitino problem were based on the gravitino number density given by Ellis et al [11]. Therefore our constraints are more stringent than others. Furthermore, our photon spectrum is different from that in Ref. [16] as shown in Fig.2. The spectrum adopted by Ref. [16] has more power to destroy light elements above threshold for the photon-photon scattering and less power below the threshold. It is expected that the difference comes mainly from the neglect of Compton scattering in Ref. [16]. In addition, the absolute value of our spectrum slightly depends on the energy of initial photons. This can be explained in the following way. If the injected photons become more energetic, they induces larger number of scattering events before thermalized, which produce larger number of secondary high energy photons. Therefore, the energy of the initial photon gets larger, the number density of high energy photons becomes higher. This effect was not taken into account in Ref. [16].

Therefore we believe that the spectrum that we have obtained is more precise than those used in other works.

In summary, we have investigated the photo-dissociation processes of light elements due to the high energy photons emitted in the decay of gravitino and set the upperbound of the reheating temperature by using precise production rate of gravitino and the spectrum of high energy photon. Together with other constraints ( the present mass density of photino and enhancement of cosmic expansion due to gravitino) we have obtained the following constraint;

$$T_R \lesssim 10^{6-7} \text{GeV} \qquad m_{3/2} \lesssim 100 \text{GeV}, \tag{28}$$

$$T_R \lesssim 10^{7-9} \text{GeV} \qquad 100 \text{GeV} \lesssim m_{3/2} \lesssim 1 \text{TeV}, \tag{29}$$

$$T_R \lesssim 10^{9-12} \text{GeV} \qquad 1 \text{TeV} \lesssim m_{3/2} \lesssim 3 \text{TeV}, \tag{30}$$

$$T_R \lesssim 10^{12} \text{GeV} \qquad 3 \text{TeV} \lesssim m_{3/2} \lesssim 10 \text{TeV}. \tag{31}$$

This provides a severe constraint in building the inflation models based on supergravity. In this paper we also study the gravitino which decays into other channels by taking the branching ratio as a free parameter. Although this gives conservative upperbound for the reheating temperature, the precise constraints cannot obtained unless various processes induced by other decay products are fully taken into account. This will be done in the future work [35].

**Acknowledgement**

We would like to thank T. Yanagida for useful comments, discussions and reading of this manuscript, and to K. Maruyama for informing us of the experimental data for photo-dissociation processes. One of the author (T.M.) thanks Institute for Cosmic Ray Research where part of this work has done. This work is supported in part by the Japan Society for the Promotion of Science.

**APPENDIX A: INTERACTION OF GRAVITINO**

In this appendix, we will discuss the interaction of the gravitino with the ordinary particles. From the supergravity lagrangian [5], we can obtain the relevant interaction terms of gravitino $\psi_\mu$ with gauge multiplets $(A_\mu, \lambda)$ and chiral multiplets $(\phi, \chi)$;

$$\mathcal{L} = \frac{i}{8M} \bar{\lambda} \gamma_\mu \left[ \gamma_\nu \gamma_\rho \right] \psi_\mu F_{\nu\rho} + \left\{ \frac{1}{\sqrt{2}M} \bar{\psi}_{\mu L} \gamma_\nu \gamma_\mu \chi_L D_\nu \phi^\dagger + h.c. \right\}. \tag{A1}$$

Note that other interaction terms including gravitino field are not important for our analysis since their contributions are suppressed by factor $M^{-1}$.

Combining Eq.(A1) with the renormalizable part of the SUSY lagrangian, we have calculated the helicity $\pm\frac{3}{2}$ gravitino production cross sections and the results are shown in Table II. Note that the cross sections for the processes (B), (F), (G) and (H) are singular because of the $t$-channel exchange of gauge bosons. These singularities should be cut off when the effective gauge boson mass $m_{eff}$ due to the plasma effect has been taken into account. Following Ref. [11], we take $\delta \equiv (1 \mp \cos\theta)_{min} = (m_{eff}^2/2T^2)$ where $\theta$ is a scattering angle in the center of mass frame, and in our numerical calculations, we have chosen $\log(m_{eff}^2/T^2) = 0$.

The total cross section in thermal bath $\Sigma_{tot}$ is defined by

$$\Sigma_{tot} = \frac{1}{2} \sum_{x,y,z} \eta_x \eta_y \, \sigma_{(x+y \to \psi_\mu + z)} \,, \tag{A2}$$

where $\sigma_{(x+y \to \psi_\mu + z)}$ is the cross section for the process $x + y \to \psi_\mu + z$, $\eta_x = 1$ for incoming bosons, $\eta_x = \frac{3}{4}$ for fermions. For the MSSM particle content, $\Sigma_{tot}$ is given by

$$\Sigma_{tot} = \frac{1}{M^2} \left\{ 2.50 g_1^2(T) + 4.99 g_2^2(T) + 11.78 g_3^2(T) \right\}, \tag{A3}$$

where $g_1$, $g_2$ and $g_3$ are the gauge coupling constants of the gauge group $U(1)_Y$, $SU(2)_L$ and $SU(3)_C$, respectively. Note that in high energy scattering processes, effect of the renormalization group flow of the gauge coupling constants should be considered. Using the one loop $\beta$-function of MSSM, solution of the renormalization group equation of gauge coupling constants is given by

$$g_i(T) \simeq \left\{ g_i^{-2}(M_Z) - \frac{b_i}{8\pi^2} \log\left(\frac{T}{M_Z}\right) \right\}^{-1/2}, \tag{A4}$$

with $b_1 = 11$, $b_2 = 1$, $b_3 = -3$. In this paper, we have used the above formula.

From Eq.(A1), we can also get the decay rate of the gravitino. In this paper, we only consider $\psi_\mu \to \gamma + \tilde{\gamma}$ decay mode, for which the decay rate is given by

$$\Gamma = \frac{m_{3/2}^3}{32\pi M^2} \left\{ 1 - \left(\frac{m_{\tilde{\gamma}}}{m_{3/2}}\right)^2 \right\}^3 \left\{ 1 + \frac{1}{3}\left(\frac{m_{\tilde{\gamma}}}{m_{3/2}}\right)^2 \right\}, \tag{A5}$$

where $m_{\tilde{\gamma}}$ is the photino mass. In the case of $m_{\tilde{\gamma}} \ll m_{3/2}$, this decay rate corresponds to the lifetime

$$\tau(\psi_\mu \to \gamma + \tilde{\gamma}) = 3.9 \times 10^8 \left(\frac{m_{3/2}}{100\,\text{GeV}}\right)^{-3} \quad \text{sec}. \tag{A6}$$

## APPENDIX B: BOLTZMANN EQUATION

In order to calculate the high energy photon spectrum, we must estimate the cascade processes induced by the radiative decay of the gravitinos. In our calculation, we have taken the following processes into account; (I) double photon pair creation, (II) photon-photon scattering, (III) pair creation in nuclei, (IV) Compton scattering off thermal electron, (V) inverse Compton scattering off background photon, and (VI) radiative decay of the gravitinos. The Boltzmann equations for this cascade processes are given by

$$\frac{\partial f_\gamma(\epsilon_\gamma)}{\partial t} = \left.\frac{\partial f_\gamma(\epsilon_\gamma)}{\partial t}\right|_{\rm DP} + \left.\frac{\partial f_\gamma(\epsilon_\gamma)}{\partial t}\right|_{\rm PP} + \left.\frac{\partial f_\gamma(\epsilon_\gamma)}{\partial t}\right|_{\rm PC}$$
$$+ \left.\frac{\partial f_\gamma(\epsilon_\gamma)}{\partial t}\right|_{\rm CS} + \left.\frac{\partial f_\gamma(\epsilon_\gamma)}{\partial t}\right|_{\rm IC} + \left.\frac{\partial f_\gamma(\epsilon_\gamma)}{\partial t}\right|_{\rm DE}, \tag{B1}$$

$$\frac{\partial f_e(E_e)}{\partial t} = \left.\frac{\partial f_e(E_e)}{\partial t}\right|_{\rm DP} + \left.\frac{\partial f_e(E_e)}{\partial t}\right|_{\rm PC} + \left.\frac{\partial f_e(E_e)}{\partial t}\right|_{\rm CS} + \left.\frac{\partial f_e(E_e)}{\partial t}\right|_{\rm IC}, \tag{B2}$$

Below, we see contributions from each processes in detail.

### (I) DOUBLE PHOTON PAIR CREATION [ $\gamma + \gamma \to e^+ + e^-$ ]

For the high energy photon whose energy is larger than $\sim m_e^2/22T$, double photon pair creation is the most dominant process.

The total cross section for the double photon pair creation process $\sigma_{DP}$ is given by

$$\sigma_{DP}(\beta) = \frac{1}{2}\pi r_e^2 \left(1 - \beta^2\right) \left\{ \left(3 - 4\beta^2\right) \log \frac{1+\beta}{1-\beta} - \beta \left(2 - \beta^2\right) \right\}, \tag{B3}$$

where $r_e = \alpha/m_e$ is the classical radius of electron and $\beta$ is the electron ( or positron ) velocity in the center of mass frame. Using this formula, one can write down $(\partial f_\gamma/\partial t)|_{\rm DP}$ as

$$\left.\frac{\partial f_\gamma(\epsilon_\gamma)}{\partial t}\right|_{\rm IC} = -\frac{2m_e^4}{\epsilon_\gamma^2} f_\gamma(\epsilon_\gamma) \int_{m_e^2/\epsilon_\gamma}^{\infty} d\bar{\epsilon}_\gamma \frac{\bar{f}_\gamma(\bar{\epsilon}_\gamma)}{\bar{\epsilon}_\gamma^2} \int_1^{\epsilon_\gamma \bar{\epsilon}_\gamma/m_e^2} ds\ \sigma_{DP}(\beta) \bigg|_{\beta=\sqrt{1-(1/s)}}. \tag{B4}$$

The spectrum of the final state electron and positron is obtained in Ref. [30], and $(\partial f_e/\partial t)|_{\rm DP}$ is given by

$$\left.\frac{\partial f_e(E_e)}{\partial t}\right|_{\rm DP} = \frac{1}{4}\pi r_e^2 m_e^4 \int_{E_e}^{\infty} d\epsilon_\gamma \frac{f_\gamma(\epsilon_\gamma)}{\epsilon_\gamma^3} \int_0^{\infty} d\bar{\epsilon}_\gamma \frac{\bar{f}_\gamma(\bar{\epsilon}_\gamma)}{\bar{\epsilon}_\gamma^2} G(E_e, \epsilon_\gamma, \bar{\epsilon}_\gamma), \tag{B5}$$

where $\bar{f}_\gamma$ represents the distribution function of the background photon at temperature $T$,

$$\bar{f}_\gamma(\bar{\epsilon}_\gamma) = \frac{\bar{\epsilon}_\gamma^2}{\pi^2} \times \frac{1}{\exp(\bar{\epsilon}_\gamma/T) - 1}, \tag{B6}$$

and function $G(E_e, \epsilon_\gamma, \bar{\epsilon}_\gamma)$ is given by

$$G(E_e, \epsilon_\gamma, \bar{\epsilon}_\gamma) = \frac{4\left(E_e + E_e'\right)^2}{E_e E_e'} \log \frac{4\bar{\epsilon}_\gamma E_e E_e'}{m_e^2(E_e + E_e')} - 8\frac{\bar{\epsilon}_\gamma \epsilon_\gamma}{m_e^2}$$
$$+ \frac{2\left\{2\bar{\epsilon}_\gamma\left(E_e + E_e'\right) - m_e^2\right\}\left(E_e + E_e'\right)^2}{m_e^2 E_e E_e'}$$
$$- \left\{1 - \frac{m_e^2}{\bar{\epsilon}_\gamma\left(E_e + E_e'\right)}\right\} \frac{\left(E_e + E_e'\right)^4}{E_e^2 E_e'^{\,2}}, \tag{B7}$$

with

$$E_e' = \epsilon_\gamma + \bar{\epsilon}_\gamma - E_e.$$

## (II) PHOTON-PHOTON SCATTERING [ $\gamma + \gamma \to \gamma + \gamma$ ]

If the photon energy is below the effective threshold of the double photon pair creation, photon-photon scattering process becomes significant. This process is analyzed in Ref. [19] and for $\epsilon'_\gamma \lesssim O(m_e^2/T)$, $(\partial f_\gamma/\partial t)|_{\mathrm{PP}}$ is given by

$$\left.\frac{\partial f_\gamma(\epsilon'_\gamma)}{\partial t}\right|_{\mathrm{PP}} = \frac{35584}{10125\pi}\alpha r_e^2 m_e^{-6} \int_{\epsilon'_\gamma}^{\infty} d\epsilon_\gamma f_\gamma(\epsilon_\gamma)\epsilon_\gamma^2 \left\{1 - \frac{\epsilon'_\gamma}{\epsilon_\gamma} + \left(\frac{\epsilon'_\gamma}{\epsilon_\gamma}\right)^2\right\}^2 \int_0^\infty d\bar{\epsilon}_\gamma \bar{\epsilon}_\gamma^3 \bar{f}_\gamma(\bar{\epsilon}_\gamma)$$
$$- \frac{1946}{10125\pi} f_\gamma(\epsilon'_\gamma)\alpha r_e^2 m_e^{-6} {\epsilon'_\gamma}^3 \int_0^\infty d\bar{\epsilon}_\gamma \bar{\epsilon}_\gamma^3 \bar{f}_\gamma(\bar{\epsilon}_\gamma). \tag{B8}$$

For a larger value of $\epsilon'_\gamma$, we cannot use this formula. But in this energy region, photon-photon scattering is not significant because double photon pair creation determines the shape of the photon spectrum. Therefore, instead of using the exact formula, we have taken $m_e^2/T$ as a cutoff scale of $(\partial f_\gamma/\partial t)|_{\mathrm{PP}}$, i.e., for $\epsilon'_\gamma \leq m_e^2/T$ we have used Eq.(B8) and for $\epsilon'_\gamma > m_e^2/T$ we have taken

$$\left.\frac{\partial f_\gamma(\epsilon'_\gamma > m_e^2/T)}{\partial t}\right|_{\mathrm{PP}} = 0. \tag{B9}$$

Note that we have checked the cutoff dependence of spectra is negligible.

## (III) PAIR CREATION IN NUCLEI [ $\gamma + N \to e^+ + e^-$ ]

Scattering off the electric field around nucleon, high energy photon can produce electron positron pair if the photon energy is larger than $2m_e$. Denoting total cross section of this process $\sigma_{PC}$, $(\partial f_\gamma/\partial t)|_{\mathrm{NP}}$ is given by

$$\left.\frac{\partial f_\gamma(\epsilon_\gamma)}{\partial t}\right|_{\mathrm{NP}} = -n_N \sigma_{PC} f_\gamma(\epsilon_\gamma), \tag{B10}$$

where $n_N$ is the nucleon number density. For $\sigma_{PC}$, we have used the approximate formula derived by Maximon [31].

Differential cross section for this process $d\sigma_{PC}/dE_e$ is given in Ref. [32], and $(\partial f_e/\partial t)|_{\mathrm{NP}}$ is given by

$$\left.\frac{\partial f_e(E_e)}{\partial t}\right|_{\mathrm{NP}} = n_N \int_{E_e+m_e}^{\infty} d\epsilon_\gamma \frac{d\sigma_{PC}}{dE_e} f_\gamma(\epsilon_\gamma). \tag{B11}$$

## (IV) COMPTON SCATTERING [ $\gamma + e^- \to \gamma + e^-$ ]

Compton scattering is one of the processes by which high energy photons lose their energy. Since the photo-dissociation of light elements occurs when the temperature drops below $\sim 0.1\mathrm{MeV}$, we can consider the thermal electrons to be almost at rest. Using the total and differential cross section at the electron rest frame $\sigma_{CS}$ and $d\sigma_{CS}/dE_e$, one can derive

$$\left.\frac{\partial f_\gamma(\epsilon'_\gamma)}{\partial t}\right|_{\rm CS} = \bar{n}_e \int_{\epsilon'_\gamma}^{\infty} f_\gamma(\epsilon'_\gamma) \frac{d\sigma_{CS}(\epsilon'_\gamma, \epsilon_\gamma)}{dE_e} - \bar{n}_e \sigma_{CS} f_\gamma(\epsilon'_\gamma), \tag{B12}$$

$$\left.\frac{\partial f_e(E'_e)}{\partial t}\right|_{\rm CS} = \bar{n}_e \int_{E'_e}^{\infty} f_\gamma(\epsilon'_\gamma) \frac{d\sigma_{CS}(\epsilon_\gamma + m_e - E'_e, \epsilon_\gamma)}{dE_e}, \tag{B13}$$

where $\bar{n}_e$ is the number density of thermal electron.

### (V) INVERSE COMPTON SCATTERING [ $e^\pm + \gamma \to e^\pm + \gamma$ ]

Contribution from the inverse Compton process is given by Jones [33], and $(\partial f/\partial t)|_{\rm IC}$ is given by

$$\left.\frac{\partial f_\gamma(\epsilon_\gamma)}{\partial t}\right|_{\rm IC} = 2\pi r_e^2 m_e^2 \int_{\epsilon_\gamma + m_e}^{\infty} dE_e \frac{2 f_e(E_e)}{E_e^2} \int_0^{\infty} d\bar{\epsilon}_\gamma \frac{\bar{f}_\gamma(\bar{\epsilon}_\gamma)}{\bar{\epsilon}_\gamma} F(\epsilon_\gamma, E_e, \bar{\epsilon}_\gamma), \tag{B14}$$

$$\left.\frac{\partial f_e(E'_e)}{\partial t}\right|_{\rm IC} = 2\pi r_e^2 m_e^2 \int_{E'_e}^{\infty} dE_e \frac{f_e(E_e)}{E_e^2} \int_0^{\infty} d\bar{\epsilon}_\gamma \frac{\bar{f}_\gamma(\bar{\epsilon}_\gamma)}{\bar{\epsilon}_\gamma} F(E_e + \bar{\epsilon}_\gamma - E'_e, E_e, \bar{\epsilon}_\gamma)$$

$$- 2\pi r_e^2 m_e^2 \frac{f_e(E'_e)}{E'^{\,2}_e} \int_{E'_e}^{\infty} d\epsilon_\gamma \int_0^{\infty} d\bar{\epsilon}_\gamma \frac{\bar{f}_\gamma(\bar{\epsilon}_\gamma)}{\bar{\epsilon}_\gamma} F(\epsilon_\gamma, E'_e, \bar{\epsilon}_\gamma), \tag{B15}$$

where function $F(\epsilon_\gamma, E_e, \bar{\epsilon}_\gamma)$ is given by

$$F(\epsilon_\gamma, E_e, \bar{\epsilon}_\gamma) = 2q \log q + (1 + 2q)(1 - q) + \frac{(\Gamma_\epsilon q)^2}{2(1 - \Gamma_\epsilon q)}(1 - q) \tag{B16}$$

with

$$\Gamma_\epsilon = \frac{4 \bar{\epsilon}_\gamma E_e}{m_e^2}, \qquad q = \frac{\epsilon_\gamma}{\Gamma_\epsilon(E_e - \epsilon_\gamma)}.$$

### (VI) GRAVITINO RADIATIVE DECAY [ $\psi_\mu \to \gamma + \tilde{\gamma}$ ]

Source of the non-thermal photon and electron spectra is radiative decay of gravitino. Since gravitinos are almost at rest when they decay and we only consider two body decay process, incoming high energy photons have unique energy $\epsilon_{\gamma 0}$, which is given by

$$\epsilon_{\gamma 0} = \frac{m_{3/2}^2 - m_{\tilde{\gamma}}^2}{2 m_{3/2}}. \tag{B17}$$

Therefore, $(\partial f_\gamma/\partial t)|_{\rm DE}$ can be written as

$$\left.\frac{\partial f_\gamma(\epsilon_\gamma)}{\partial t}\right|_{\rm DP} = \delta(\epsilon_\gamma - \epsilon_{\gamma 0}) \frac{n_{3/2}}{\tau_{3/2}}. \tag{B18}$$

FIGURES

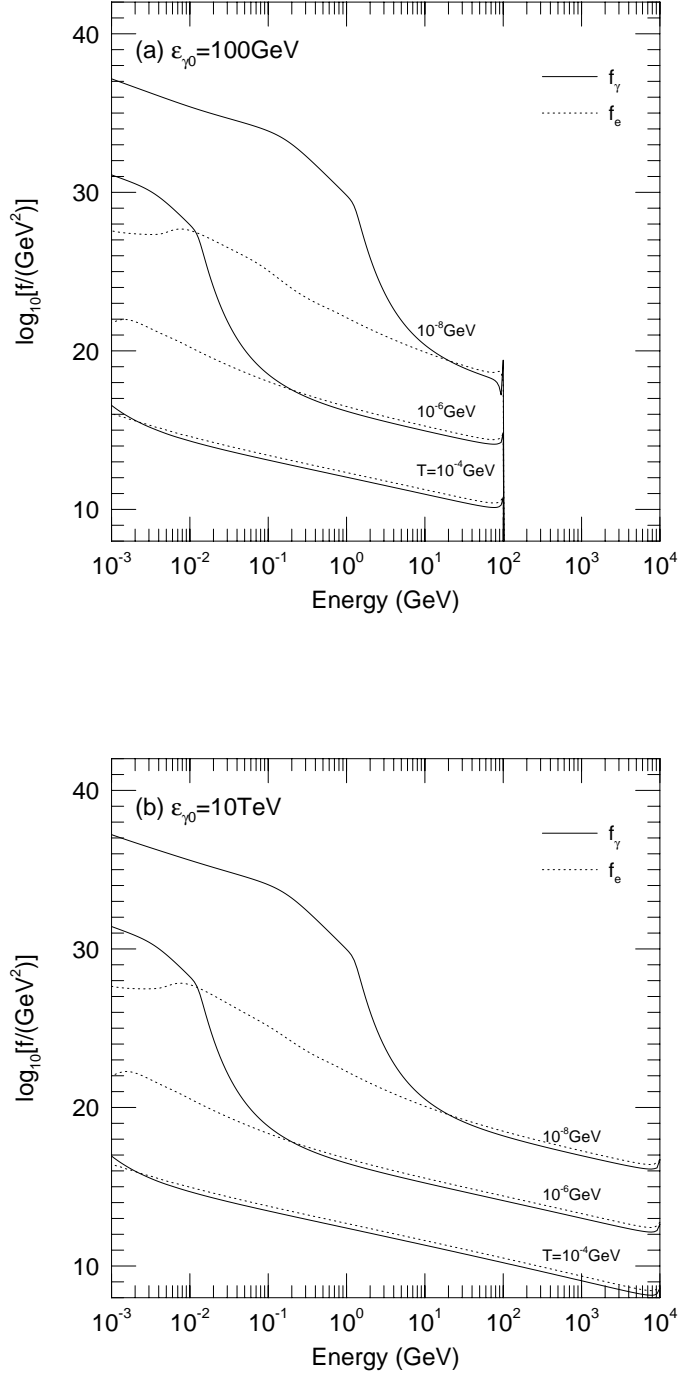

FIG. 1. Typical spectra of photon (the solid lines) and electron (the dotted lines). We have taken the temperature of the background photon to be $T = 100\text{keV}, 1\text{keV}, 10\text{eV}$, and the energy of the incoming high energy photon $\epsilon_{\gamma 0}$ is (a) 100GeV and (b) 10TeV. Normalization of the initial photon is given by $\epsilon_{\gamma 0} \times (\partial \tilde{f}_\gamma(\epsilon_\gamma)/\partial t)|_{\text{DE}} = \delta(\epsilon_\gamma - \epsilon_{\gamma 0})$ GeV$^5$.

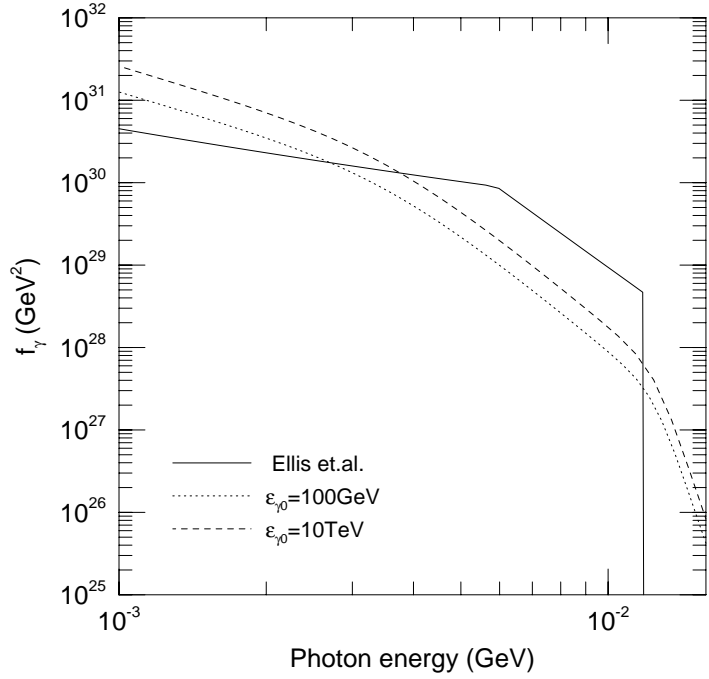

FIG. 2. Photon spectrum derived from the fitting formula used in Ref. [16] is compared with our results. We have taken the temperature of the background photon to be 100eV and the normalization of the incoming flux is the same as Fig.1. The solid line is the result of fitting formula, and the rest of two are our results with $\epsilon_\gamma = 100\text{GeV}$ (dotted line) and $\epsilon_\gamma = 10\text{TeV}$ (dashed line).

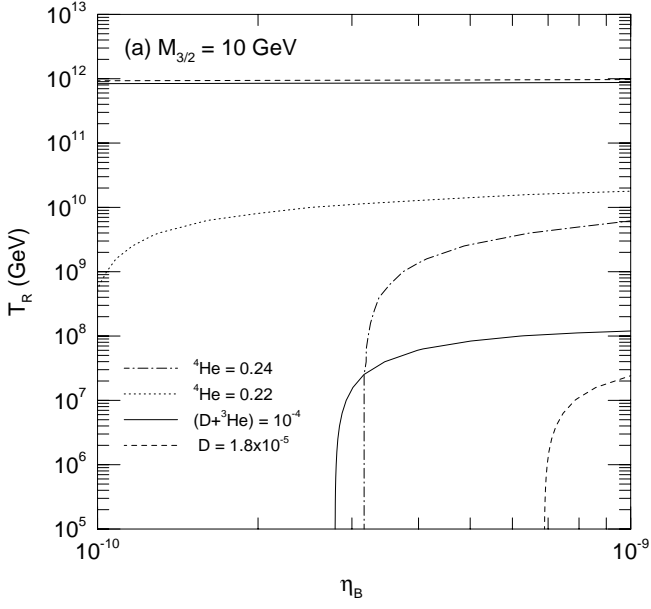 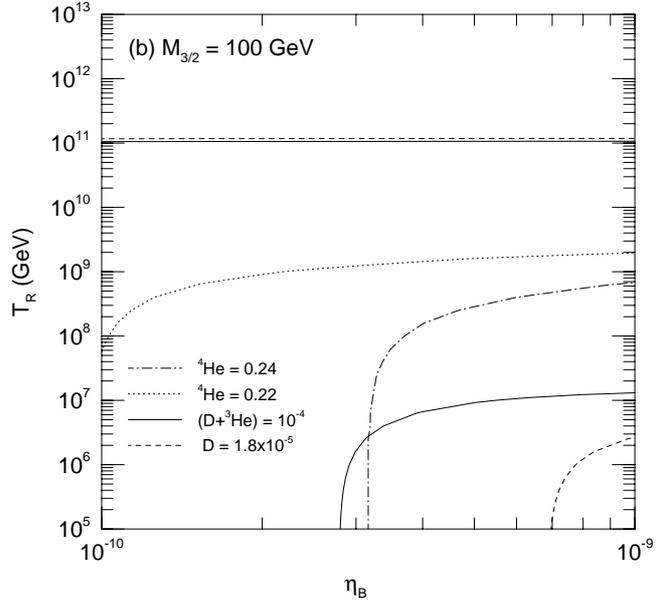
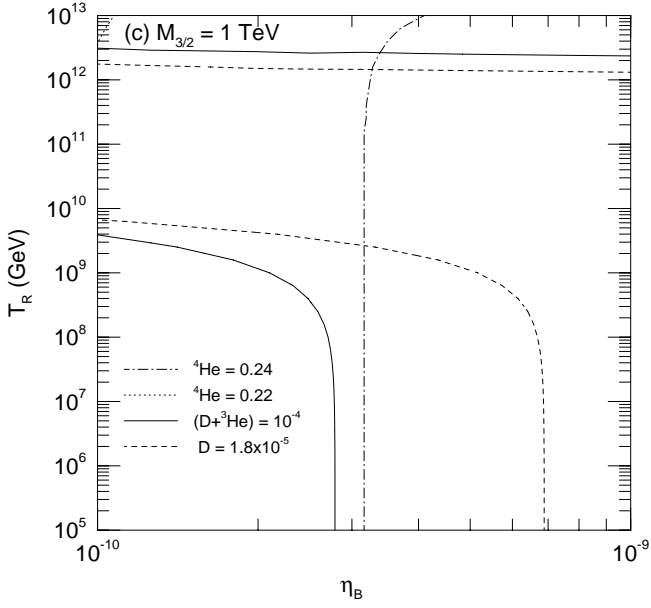 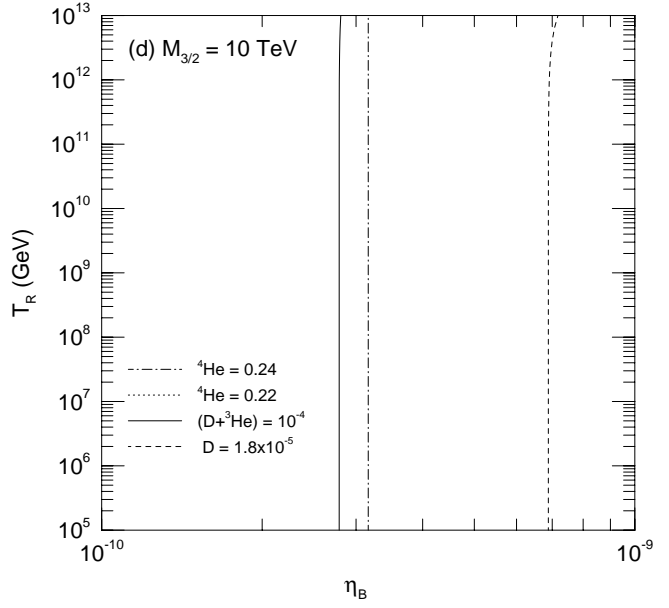

FIG. 3. Contours for critical abundance of light elements in the $\eta_B - T_R$ plane for (a) $m_{3/2} = 10$GeV, (b) $m_{3/2} = 100$GeV, (c) $m_{3/2} = 1$TeV and (d) $m_{3/2} = 10$TeV.

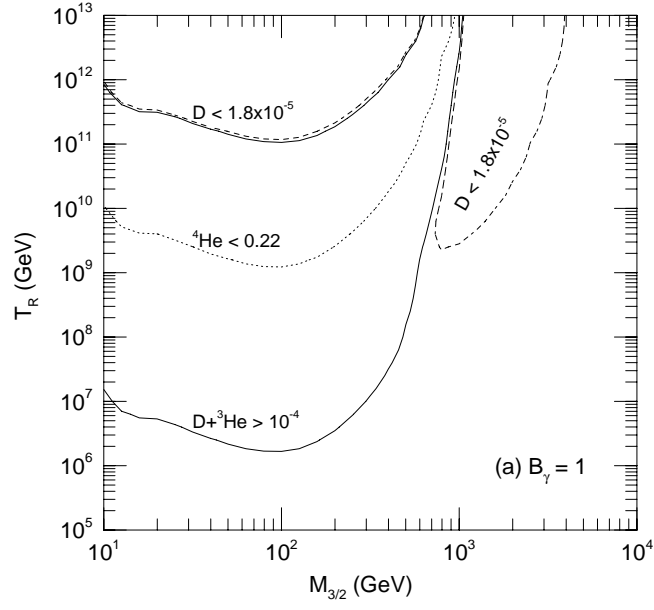

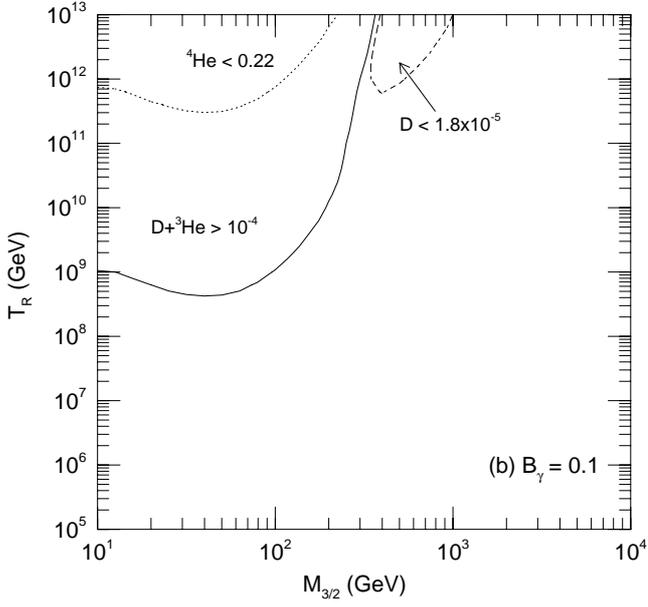

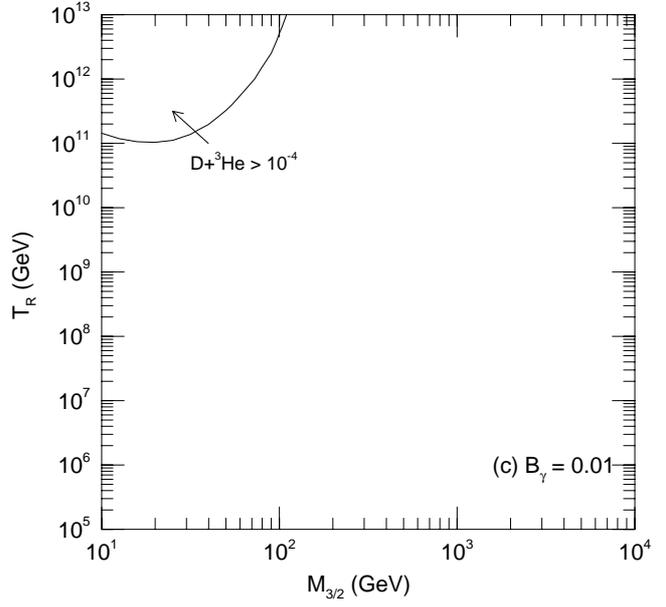

FIG. 4. Allowed regions in $m_{3/2} - T_R$ plane for (a) $B_\gamma = 1$, (b) $B_\gamma = 0.1$ and (c) $B_\gamma = 0.01$. In the region above the solid curve $^3$He and D are overproduced, the abundance of $^4$He is less than 0.22 above the dotted curve and the abundance of D is less than $1.8 \times 10^{-5}$ above the dashed curve.

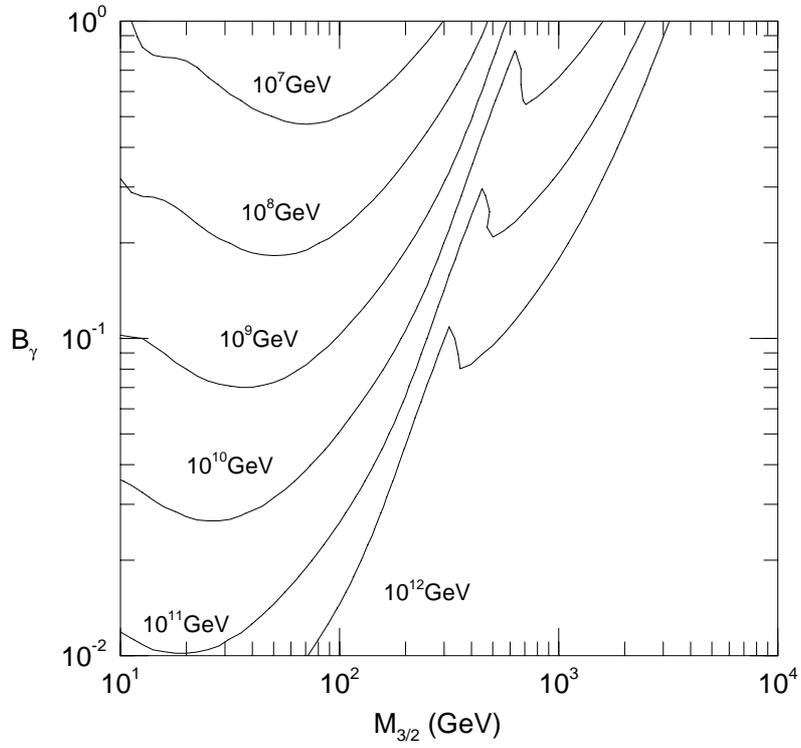

FIG. 5. Contours for the upper limits of the reheating temperature in the $m_{3/2} - B_\gamma$ plane. The numbers in the figure denote the limit of the reheating temperature.

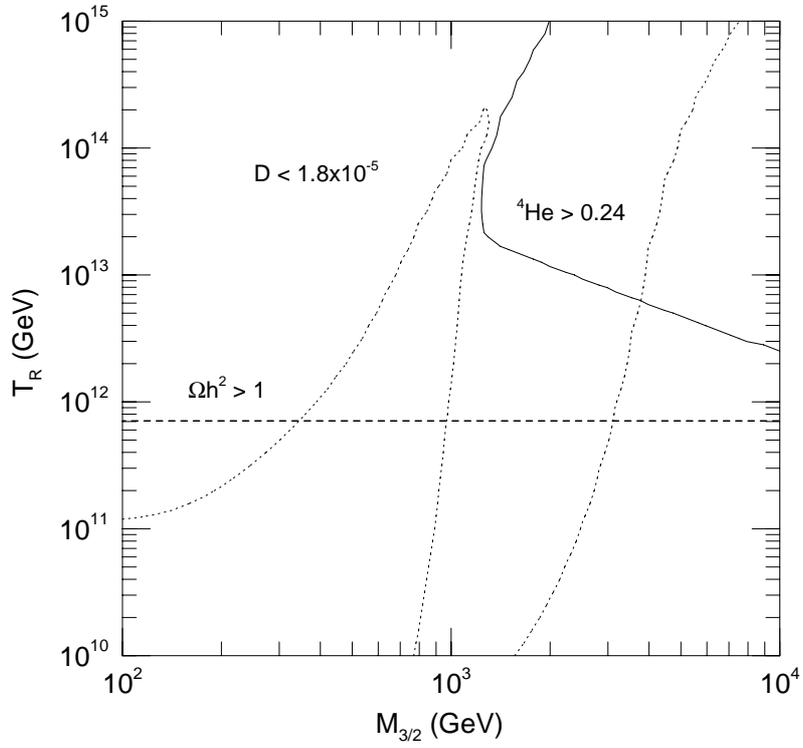

FIG. 6. Upperbound of the reheating temperature. Dashed line represents the constraint from the present mass density of photino. Solid curve represents the upperbound requiring $^4$He < 0.24. Constraints from D photo-dissociation is also shown by dotted line.

TABLES

| Reaction | Threshold (MeV) | References |
|---|---|---|
| D $+\gamma \to n + p$ | 2.225 | [21] |
| T $+\gamma \to n+$ D | 6.257 | [22], [23] |
| T $+\gamma \to p + n + n$ | 8.482 | [23] |
| $^3$He $+\gamma \to p+$D | 5.494 | [24] |
| $^3$He $+\gamma \to p+$D | 7.718 | [24] |
| $^4$He $+\gamma \to p+$ T | 19.815 | [25] |
| $^4$He $+\gamma \to n+^3$He | 20.578 | [26] |
| $^4$He $+\gamma \to p + n+$ D | 26.072 | [27] |

TABLE I. Photo-disociation reactions

| Process | | $\sigma = (g^2/64\pi M^2) \times$ |
|---|---|---|
| (A) | $A^a + A^b \to \psi + \lambda^c$ | $(8/3)\left|f^{abc}\right|^2$ |
| (B) | $A^a + \lambda^b \to \psi + A^c$ | $4\left|f^{abc}\right|^2 \{-(3/2) + 2\log(2/\delta) + \delta - (1/8)\delta^2\}$ |
| (C) | $A^a + \phi_i \to \psi + \chi_j$ | $4\left|T^a_{ji}\right|^2$ |
| (D) | $A^a + \chi_i \to \psi + \phi_j$ | $2\left|T^a_{ji}\right|^2$ |
| (E) | $\chi_i + \phi^*_j \to \psi + A^a$ | $4\left|T^a_{ji}\right|^2$ |
| (F) | $\lambda^a + \lambda^b \to \psi + \lambda^c$ | $\left|f^{abc}\right|^2 \{-(62/3) + 16\log[(2-\delta)/\delta] + 22\delta - 2\delta^2 + (2/3)\delta^3\}$ |
| (G) | $\lambda^a + \chi_i \to \psi + \chi_j$ | $4\left|T^a_{ji}\right|^2 \{-2 + 2\log(2/\delta) + \delta\}$ |
| (H) | $\lambda^a + \phi_i \to \psi + \phi_j$ | $\left|T^a_{ji}\right|^2 \{-6 + 8\log(2/\delta) + 4\delta - (1/2)\delta^2\}$ |
| (I) | $\chi_i + \bar\chi_j \to \psi + \lambda^a$ | $(8/3)\left|T^a_{ji}\right|^2$ |
| (J) | $\phi_i + \phi^*_j \to \psi + \lambda^a$ | $(16/3)\left|T^a_{ji}\right|^2$ |

TABLE II. Total cross sections for the helicity $\pm\frac{3}{2}$ gravitino production process. Spins of the initial states are averaged and those of the final states are summed. $f^{abc}$ and $T^a_{ij}$ represent the structure constants and the generator of the gauge group, respectively. Note that for the processes (B), (F), (G) and (H), we cut off the singularities due to the $t$-, $u$-channel exchange of gauge bosons, taking $(1 \pm \cos\theta)_{min} = \delta$ where $\theta$ is the scattering angle in the center-of-mass frame.